# Multiple glassy dynamics of a homologous series of triphenylene-based columnar liquid crystals – A study by broadband dielectric spectroscopy and advanced calorimetry

Arda Yildirim[1,2,3,4], Christina Krause[1], Patrick Huber[2,3,4], and Andreas Schönhals[1,*]

[1]Bundesanstalt für Materialforschung und -prüfung (BAM), Unter den Eichen 87,12205 Berlin, Germany

[2]Hamburg University of Technology, Institute for Materials and X-Ray Physics, Denickestr. 10, 21073 Hamburg, Germany

[3]Centre for Hybrid Nanostructures CHyN, University Hamburg, Luruper Chaussee 149, 22761 Hamburg, Germany

[4]Centre for X-Ray and Nano Science CXNS, Deutsches Elektronen-Synchrotron DESY, Notkestr. 85, 22607 Hamburg, Germany

**TOC entry:**

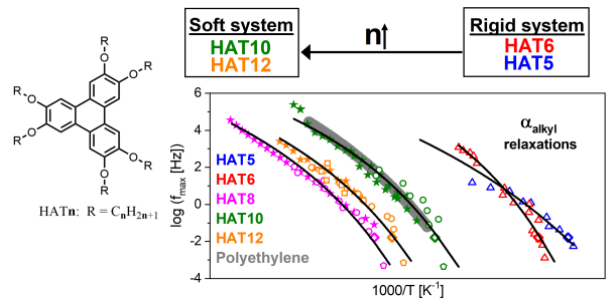

# ABSTRACT

Hexakis(n-alkyloxy)triphenylene) (HATn) consisting of an aromatic triphenylene core and alkyl side chains are model discotic liquid crystal (DLC) systems forming a columnar mesophase. In the mesophase, the molecules of HATn self-assemble in columns, which has one-dimensional high charge carrier mobility along the columns. Here, a homologous series of HATn with different length of the alkyl chain (n=5,6,8,10,12) is investigated using differential scanning calorimetry (DSC), broadband dielectric spectroscopy (BDS) and advanced calorimetric techniques including fast scanning calorimetry (FSC) and specific heat spectroscopy (SHS). The investigation of the phase behavior was done utilizing DSC experiments and the influence of the alkyl chain length on the phase behavior was revealed. By the dielectric investigations probing the molecular mobility, a γ-relaxation due to localized fluctuations as well as two glassy dynamics the $\alpha_{core}$- and $\alpha_{alkyl}$-relaxation were observed in the temperature range of the plastic crystalline phase. Moreover, the observed glassy dynamics were further studied employing advanced calorimetry. All observed relaxation processes are attributed to the possible specific molecular fluctuations and discussed in detail. From the results a transition at around n=8 from a rigid constrained (n=5,6) to a softer system (n=10,12) was revealed with increasing alkyl chain length. A counterbalance of two competing effects of a polyethylene-like behavior of the alkyl chains in the intercolumnar domains and self-organized confinement is discussed in the context of a hindered glass transition.

# 1. INTRODUCTION

Liquid crystals (LCs) forming a columnar mesophase are classified as so-called columnar liquid crystals (CLCs). They have a 1D high charge carrier mobility in the columnar mesophase and therefore they are qualified for a variety of potential applications like organic field-effect transistors, organic light-emitting diodes, and organic photovoltaics just to mention a few for the field of electronics besides others.[1,2] Since the discovery of columnar mesophases formed by a LC in the late 1970s,[3] the interest in synthesizing different CLCs and investigating their properties have grown to disclose their potential in electronic related applications and in photonics.[4-6]

For discotic liquid crystals (DLCs), the individual molecules self-assemble into different types of columnar mesophases such as rectangular, oblique, square and hexagonal columnar phases. These materials are considered as CLCs together with other LCs forming columnar mesophase such as ionic liquid crystals (ILCs) as one example.[7] Among a variety of DLCs, hexakis(n-alkyloxy)triphenylenes **HATn** (where n is the number of carbon atoms in substituents) are triphenylene-based DLCs and have been extensively investigated since they provide a relatively simple (molecularly symmetric) DLC model system. Thus, **HATn** are the one of the most studied DLC materials in the DLC research.[8]

A molecule of the **HATn** series consists of a rigid disk-like aromatic triphenylene core and flexible alkyl chains bonded to the core via ether linkages (see Figure 1a). The molecules of **HATn** self-organize into columns which are hexagonally ordered in the hexagonal columnar mesophase (Col$_h$) observed at temperatures between a plastic crystalline (Cry) at low temperatures and the isotropic (Iso) phase at higher temperatures. The unfavorable interactions between the rigid core (order) and the flexible substituents (mobility) as well as the favorable π-π interaction among the aromatic cores, cause the self-assembly into columns which can be considered as 1D liquid.[9] On the one hand, the former effect leads to nanophase separation, which has been confirmed by X-ray diffraction (XRD) for CLCs including **HATn**[9,10] as well as for different polymers possessing long n-alkyl side chains in their structure.[11-14] On the other hand, the favorable π-π interaction results in 1D high charge mobility along the column direction originating from the delocalization of π-electrons in the π-π stacked system of the **HATn** molecules in the $Col_h$ phase, where the flexible alkyl chains fill the intercolumnar area and are disordered among ordered columns of the cores (see Figure 1b).

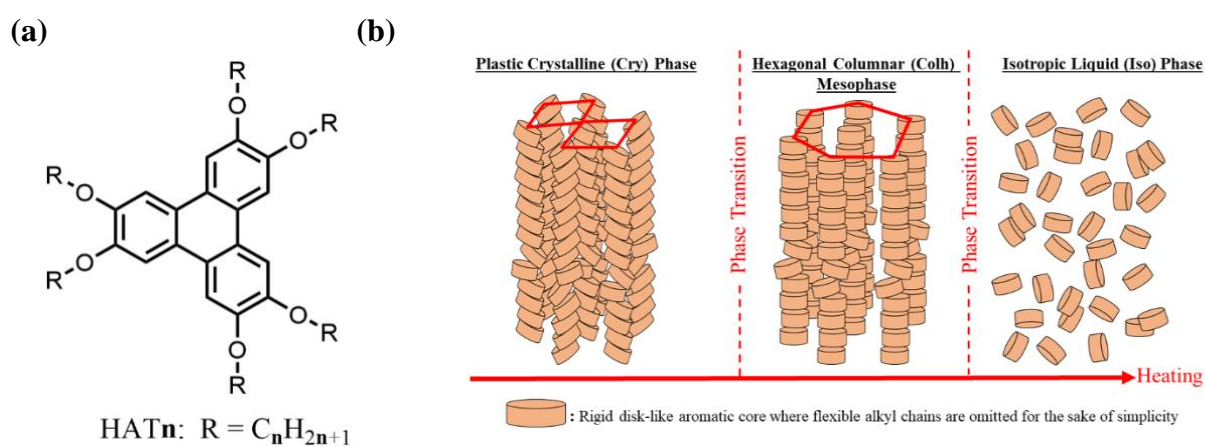

**Figure 1.** (a) Chemical structures of the triphenylene-based CLCs **HATn**. (b) Illustration of the structures formed by **HATn** molecules in the different phases.

Majority of the studies reported in the literature regarding triphenylene-based DLCs **HATn** concentrate on the synthesis, the structure, the charge carrier mobility and the phase behavior of these materials.[8,9] Among them the charge carrier mobility and phase behavior are associated with the conductivity and the stability of the columnar structure respectively, which define the application type and the corresponding service temperature range. Additionally, the molecular mobility on different timescales plays presumably a crucial role for the conductivity and stability of the columnar structure.[15,16] Therefore, besides the structure and the phase behavior the molecular dynamics of CLCs must be revealed.

The investigations about the molecular mobility of CLCs in the literature[17-23] reported multiple glassy dynamics probed in the plastic crystalline phase and/or in the columnar mesophase. The work of Suga and Seki is the first report of a glass transition investigated by calorimetry for plastic crystals.[24] After this first report about a glass transition in plastic crystals, glassy dynamics in plastic crystals are further reported and discussed in the literature.[25-29] Recently, the molecular mobility was studied for several CLC systems including **HAT6** as well as **HAT5** including dipole functionalized derivatives of the latter[30-34]. These investigations revealed a localized molecular mobility and multiple glassy dynamics detected in the Cry phase, which was also reported by other groups.[17-23] For triphenylene-based DLCs, the detected glassy dynamics are attributed to cooperative fluctuations taking place in the columns and in the intercolumnar area. The faster glassy dynamics at lower temperatures (denoted as $\alpha_2$-relaxation[31-33]) are assigned to rotational fluctuations of the disc-like core around the column axis coupled with parallel and perpendicular small scale motions of the triphenylene cores. The glassy dynamics taking place in the columns can be therefore referred as glass transition of a one-dimensional liquid. The slower glassy dynamics at higher temperatures (denoted as $\alpha_1$-relaxation[31-33]) is assigned to the polyethylene-like (**PE**-like) fluctuations of the alkyl chains in the intercolumnar area coupled with a tumbling (tilt and twist motions) of the core.

Besides these glassy dynamics, a **PE**-like localized dynamics was observed in the Cry phase for triphenylene-based DLCs and attributed to the fluctuations of methylene groups in the alkyl chains.[31-33] This **PE**-like localized dynamics takes place in the intercolumnar area, where the flexible alkyl substituents are confined between the rigid columns built by the aromatic core. Based on systematic investigations of this PE-like localized dynamics in **HATn**, the self-confinements of alky chains between relatively solid columns were shown for **HATn**.[31] Due to these studies, an understanding of the molecular mobility in CLCs and underlying mechanism has been gained. Nevertheless, a better understanding of the structure-property-molecular mobility relationships is required to tune and to improve the properties of CLCs towards their applications. Furthermore, systematic investigations of the glassy dynamics of CLCs might provide an insight into the glass transition phenomena in general to uncover this longstanding topical problem of condensed matter research.

## 2. EXPERIMENTAL SECTION

**Materials.** A homologous series of hexakis(n-alkyloxy)triphenylene) (**HATn**) DLCs consisting of an aromatic triphenylene core and side chains with different side chain length (n = 8, 10, 12) were investigated, where data for **HAT5** and **HAT6** were taken from elsewhere.[31,32] Figure 1 gives the chemical structure of the materials and a schematic illustration of their molecular organization in the Cry, Col$_h$ and Iso phase. All materials were purchased from Synthon Chemicals (Bitterfeld, Germany). They are used as they are received from the producer.

**Broadband Dielectric Spectroscopy (BDS).** The complex dielectric permittivity $\varepsilon^*(f) = \varepsilon'(f) - i\varepsilon''(f)$ was measured by isothermal frequency scans at frequencies between $10^{-1}$ Hz and $10^9$ Hz, where $\varepsilon'$ and $\varepsilon''$ are the real and imaginary (loss) part of $\varepsilon^*(f)$. $i = \sqrt{-1}$ symbolizes the imaginary unit and f denotes the frequency. For all measurements, capacitor of the samples was prepared in parallel plate geometry, where the samples were placed and kept between two gold-plated disk-shaped brass electrodes. 50 μm thick fused silica spacers were used to provide the spacing between the parallel electrodes.

In the frequency range from $10^{-1}$ Hz to $10^6$ Hz, the measurements were performed using a high-resolution Alpha analyzer with an active sample head (Novocontrol, Montabaur, Germany) in the temperature range from 163 K to 403 K. More details can be found in ref. 35. The diameter of the sample used in these measurements was 20 mm.

In the frequency range from $10^6$ Hz to $10^9$ Hz, the measurements were conducted utilizing a coaxial reflectometer based on the Agilent E4991 RF impedance analyzer in the temperature range from 173 K to 373 K. More details of the technique are given in ref. 35. For these measurements, the diameter of the sample was 6 mm.

For both measurement setups, the temperature of the sample during the measurements was regulated by a Quatro Novocontrol temperature controller facilitating nitrogen as heating and cooling agent. It provides a temperature stability of < 0.1 K.[35]

**Differential Scanning Calorimetry (DSC) and Temperature Modulated DSC (TMDSC).** For conventional DSC measurements, the sample (ca. 4 mg), placed in a 50 µl aluminum pan, was measured in the temperature range from 203 K to 423 K with a heating/cooling rate of 10 K min$^{-1}$ using a Perkin Elmer DSC 8500 instrument. Nitrogen was used as purge gas, where its flow rate was set to 20 ml min$^{-1}$. One heating cycle followed by a cooling and then by a second heating run were employed as temperature program. The baseline of the device was obtained by conducting a measurement of an empty 50 µl aluminum pan under the same conditions and subtracted from the data of the sample. The calibration of the device was checked by measuring an indium standard before the measurement.

Furthermore, TMDSC was carried out by the Perkin Elmer DSC 8500 employing the StepScan approach (StepScan DSC−SSDSC is a trademark of PerkinElmer) in order to perform specific heat spectroscopy (SHS) at a low frequency. In the SSDSC approach, alternating short heating and isothermal steps were employed. The length of the isothermal step defines the modulation frequency. The area under the heat-flow peak together with the height of the temperature step in the temperature program applied are used to estimate the StepScan specific heat capacity without applying a Fourier transformation.[36]

SSDSC measurements was conducted at a frequency of $1.6 \times 10^{-2}$ Hz (corresponding to isothermal period of 60 s) with a constant step height of 2 K and a heating rate of 80 K min$^{-1}$. Synthetic sapphire (α-Al2O3) was used as reference material to obtain absolute values of the complex specific heat capacity.

**Fast Scanning Calorimetry (FSC) and Temperature Modulated FSC (TMFSC).** FSC was carried out using a Mettler Toledo Flash DSC1 to investigate the glass transition of the samples. The chip-based power-compensated Flash DSC measurements were carried out with heating rates from 10 K s$^1$ to 10000 K s$^{-1}$.[37] The MultiSTAR UFS 1 twin chip sensor was used as sample cell.[38] Nitrogen having a flow rate of 40 ml min$^{-1}$ was used as purge gas. During the measurements, the base temperature was controlled by a Huber TC100 intercooler. Prior to measurements with a sample on a chip, the chip sensors were conditioned and corrected according to "conditioning" and "correction" procedures provided by the manufacturer.

The samples were placed onto the chip sensors and melted by heating them from room temperature with a heating rate of 100 K s$^{-1}$ to the isotropic state at 373 K. From the isotropic state, the sample is cooled down with a cooling rate of 100 K s$^{-1}$ to 293 K.

The FSC measurements were performed in the temperature range from 183 K to 313 K with heating rates from 10 K/s to 10000 K/s. Prior to every heating measurement, a cooling run having a rate of 100 K/s was applied in order to have the same thermal history for all samples before the heating run.

In addition to the static FSC measurements, SHS was carried out by FSC employing a temperature modulated approach at higher frequencies than for TMDSC. Like TMDSC, for the TMFSC measurements a step response approach applying alternating heating and isothermal steps were used.[36,39-41] A constant step height of 2 K at heating rates of 200 K s$^{-1}$ and 2000 K s$^{-1}$ with isothermal periods of 1 s and 0.1 s were applied. The ratio of the Fourier transformations of the heat flow and the instantaneous heating rate was calculated to obtain the complex heat capacity.[36,42] The temperature dependence of the modulus of the complex heat capacity, $|c_p^*| = c_{p,rev} = \sqrt{{c_p'}^2 + {c_p''}^2}$ is considered for analysis, where $c_p'$ and $c_p''$ are the real and the imaginary part of $c_p^*$. The length of the isothermal periods set prior to measurement defines the frequencies of the measurement. Moreover, different higher harmonics of the base frequencies were also analyzed.

## 3. RESULTS AND DISCUSSION

**Phase Behavior.** The phase behavior of the **HATn** compounds was investigated conducting conventional DSC experiments. The DSC thermograms for **HATn** are given in Figure 2a. As known form the literature[10], the DSC heat flow curves showed that three phases (Cry, Col$_h$ and Iso phases) are formed for all the DLCs under investigation. The observed peaks indicating phase transitions were quantitively analyzed for the second heating run. The determined phase transition temperatures and enthalpies as well as the corresponding literature values for homologous compounds **HAT5**[31] and **HAT6**[32] are depicted in

Table **1**. The estimated values are in good agreement with the reported values in the literature for **HATn**.[10,43-45]

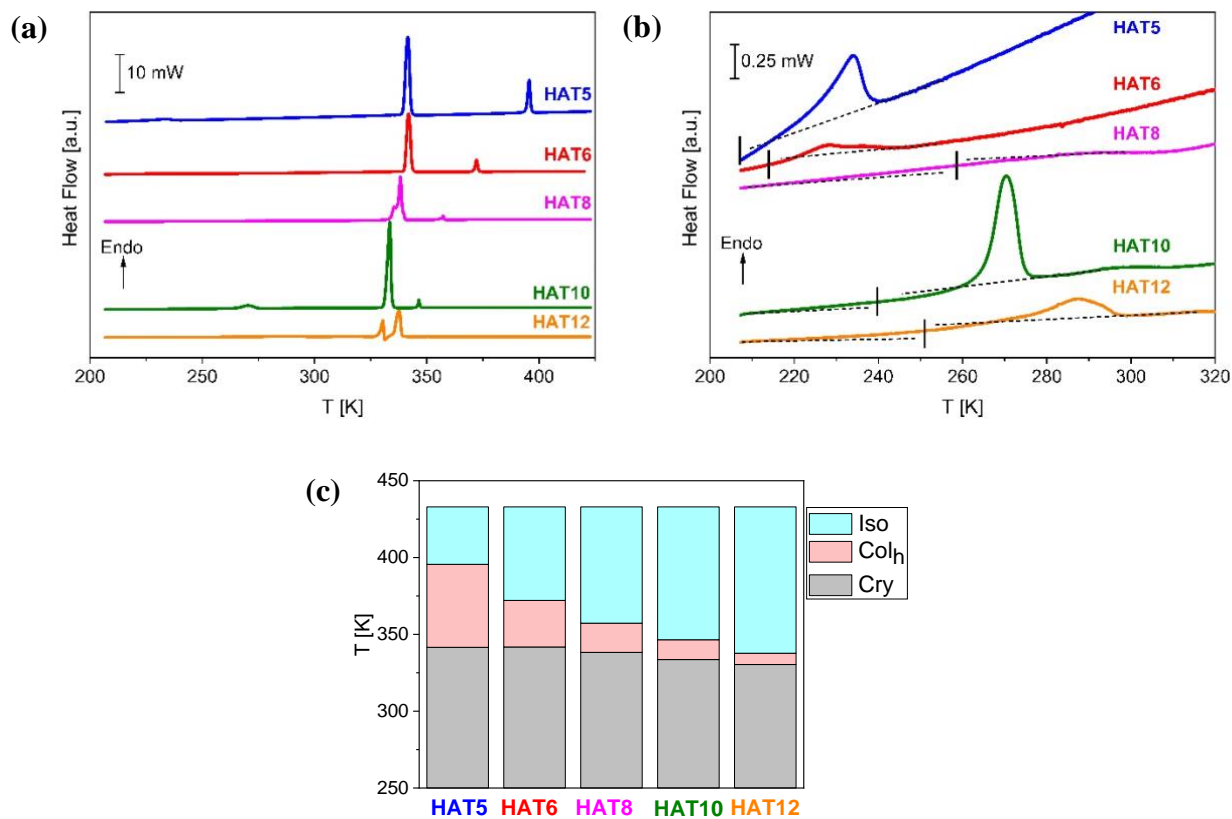

**Figure 2.** (a) DSC thermograms of the **HATn** samples for the second heating run with a heating rate of 10 K min⁻¹. (b) DSC thermograms of **HATn** zooming in the temperature range between 200 K and 320 K for the second heating run. Black solid vertical lines indicate the determined glass transition temperatures. The curves are shifted along the y-scale for the sake of clarity (shifts indicated by the mW values in the brackets). (c) Mesophase ranges of the **HATn** samples, constructed from the results given in Table 1. DSC data corresponding to **HAT5** and **HAT6** are taken from ref. 31 and ref. 32 respectively.

**Table 1.** Phase transitions temperatures and enthalpies determined by DSC

| Sample | $T_g$ [K] ($\Delta c_p$ [J K⁻¹ g⁻¹]) | Phase | T [K] ($\Delta H$ [J g⁻¹]) | Phase | T [K] ($\Delta H$ [J g⁻¹]) | Phase |
|---|---|---|---|---|---|---|
| **HAT5** from ref. 31 | 206.6 (0.15) | Cry | 341.6 (48.0) | Col$_h$ | 395.6 (13.5) | Iso |
| **HAT6** from ref. 32 | 214.0 (0.25) | Cry | 341.8 (51.3) | Col$_h$ | 372.1 (6.8) | Iso |
| **HAT8** | 257.8 (0.12) | Cry | 338.3 (82.1) | Col$_h$ | 357.3 (4.2) | Iso |
| **HAT10** | 239.1 (0.17) | Cry | 333.6 (68.3) | Col$_h$ | 346.5 (3.1) | Iso |
| **HAT12** | 251.3 (0.23) | Cry | 330.4 (21.8) | Col$_h$ | 337.7 (52.2) | Iso |

Table 1 shows that both phase transition temperatures depend on the alkyl chain length. To compare the phase transition temperatures better, a bar chart representation of the mesophase ranges is given in Figure 2c. As previously discussed[10] both the phase transition temperatures of Cry-Col$_h$ and Col$_h$-Iso transitions decrease with increasing alkyl chain length. The reduction in the phase transition temperatures is found to be more pronounced for the Col$_h$-Iso transition compared to the Cry-Col$_h$ transition. Moreover, the temperature range of the Col$_h$ mesophase becomes narrower with increasing alkyl chain length. The two major factors defining the stability of the columns are the core-core interactions and the flexibility of the alky chains. Longer alkyl chains are more flexible (less rigid) enhancing the flowability of the molecules of the DLCs which results in a decrease in the clearing (Col$_h$-Iso transition) temperature.[43] Furthermore, the packing of the cores becomes more disturbed as the length of the substituents increases. This leads to a smaller overlap of the π-orbitals of the cores and thus to a decrease of the Cry-Col$_h$ transition temperatures.[46] Therefore, this can be considered as an indirect effect of the alkyl chains on the core-core interactions defining the stability of the columnar mesophase. Moreover, a direct effect of core-core interactions on the stability of columnar mesophase have been discussed in ref. 31 for **HAT5** and its dipole-functionalized derivatives.

In the Cry phase, a step-like change was observed in the heat flow in the temperature range between 200 K and 320 K for all **HATn** compounds (see Figure 2). These steps are considered as indication of a glass transition. From the steps, a glass transition temperature (T$_g$) and the changes in the specific heat capacity (Δc$_p$) at the glass transition were estimated (see Table 1). A glass transition is also detected in Cry phase for the related CLCs in a similar temperature range than that found for **HATn**.[17-19,47] As discussed in ref. 31 for **HAT5** and its dipole functionalized derivatives as well as for **HAT6** in ref. 32, a nanophase separation of the ordered solid aromatic cores and the disordered flexible alkyl chains surrounding the aromatic core induces a disorder in the system, which is assumed to be one reason responsible for a glass transition in such materials. Moreover, it is discussed that the glass transition in such systems corresponds to the glassy dynamics of the alkyl chains.[31-33] X-ray scattering investigations evidenced a nanophase separation for triphenylene-based DLCs.[10]

**Molecular Mobility.** Error! Reference source not found. presents the dielectric spectra of the **HATn** compounds for n=8,10, and 12 in 3D representations, where dielectric loss is plotted in dependence on frequency and temperature. Three dielectrically active processes are observed as peaks in the temperature range of the Cry phase. The γ-process appears at lowest temperatures (highest frequency). At higher temperatures the $\alpha_{core}$-process is detected while the $\alpha_{alkyl}$-process appears at even higher temperatures than the other two processes. It should be noted that the $\alpha_{core}$-process is not observed for **HAT8** because the data of **HAT8** have higher errors which might make it more difficult to observe and to analyze it (see **Error! Reference source not found.**a). In this connection, it is interesting to note that **HAT8** also shows an anomalous behavior in the low frequency vibrational density of states investigated by inelastic neutron scattering.[10]

(a)

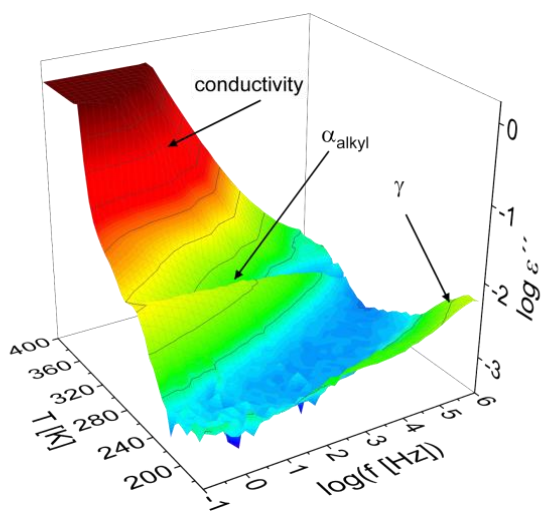

(b)

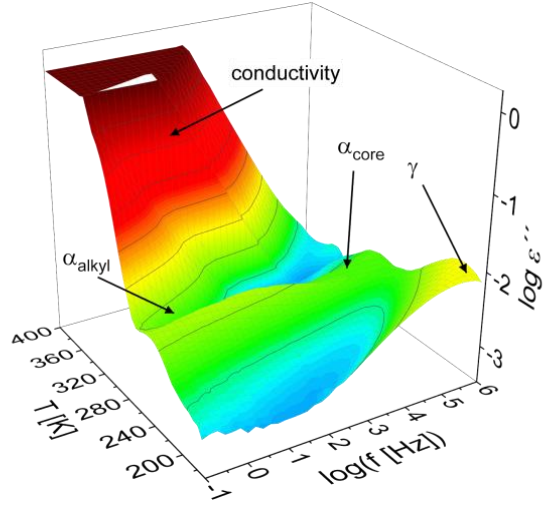

(c)

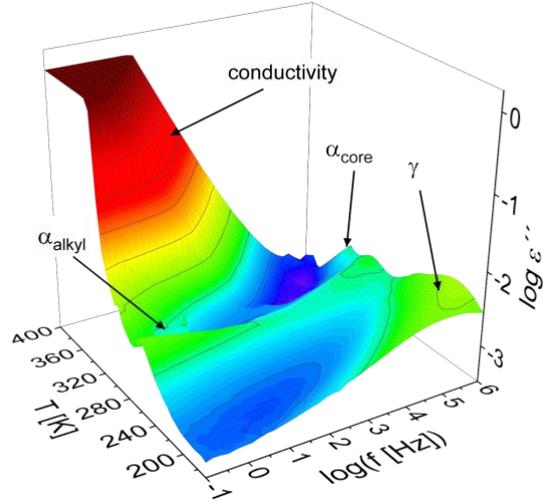

**Figure 3.** Dielectric loss as function of frequency and temperature given in 3D representations for (a) **HAT8**, (b) **HAT10** and (c) **HAT12**. Arrows show the detected dielectric processes and the conductivity contribution observed.

The relaxation processes were quantitatively analyzed by fitting the empirical Havriliak/Negami function (HN-function) to the data, which is generally used to describe non-Debye-like relaxation processes. The HN-function reads [48]

$$\varepsilon^*(\omega) = \varepsilon'(\omega) - i\varepsilon''(\omega) = \varepsilon_\infty + \frac{\Delta\varepsilon}{(1 + (i\omega\tau_{HN})^\beta)^\gamma} \quad (1)$$

where the fractional shape parameters $\beta$ and $\gamma$ ($0 < \beta; \beta\gamma \leq 1$) describe the symmetric and asymmetric broadening of the relaxation spectrum with respect to that of the Debye function. $\varepsilon_\infty$ denotes $\varepsilon'$ in the limit $\varepsilon_\infty = \lim_{\omega\to\infty}\varepsilon'(\omega)$ and $\Delta\varepsilon_{HN}$ is the dielectric strength. $\tau_{HN}$

represents the relaxation time corresponding to the frequency of maximal dielectric loss $f_{max}$. If more than one relaxation process is observed in the considered frequency window a sum of HN-functions is fitted to the data. For details see ref. 49.

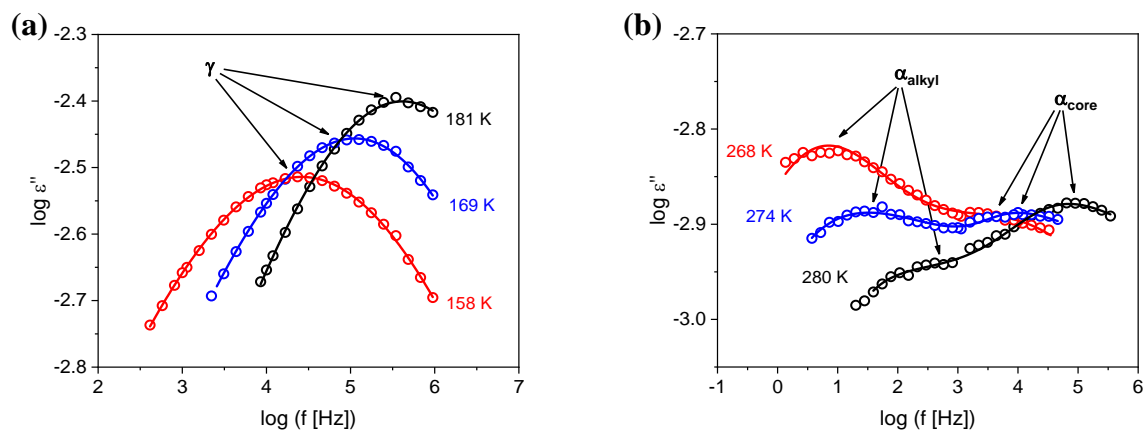

**Figure 4.** Frequency dependence of the dielectric loss of (a) the γ-process and (b) the $α_{core}$- and $α_{alkyl}$- processes of **HAT12** at the indicated temperatures. Lines are HN-fits to the data.

Figure 4 gives examples of the fitting of the HN-function to dielectric loss data of **HAT12**. The obtained $f_{max}$ values were plotted versus inverse temperature in Arrhenius coordinates (relaxation map) for all **HATn** compounds including **HAT5**[31] and **HAT6**[32] in Figure 5 and Figure 6. Additionally, the relaxation rates of all detected processes are given together in a single relaxation map in Figure S1 in the SI.

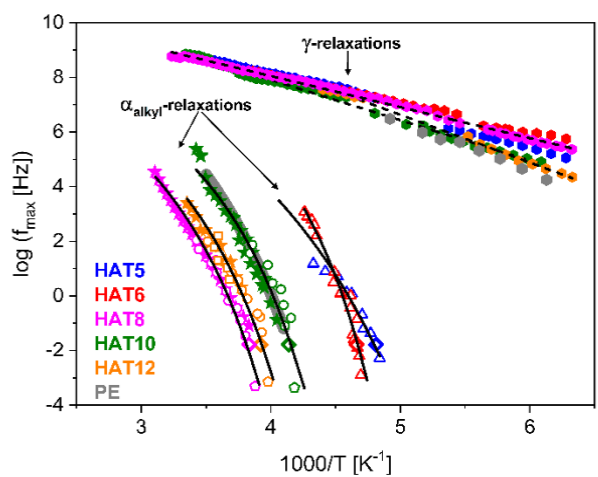

**Figure 5.** Relaxation map of **HATn** for the γ- and the $α_{alkyl}$-relaxation. Each **HATn** material is represented by a different color as indicated. Filled symbols denote the dielectric and open symbols denote the calorimetric data. Filled

hexagons – γ-relaxation (BDS); filled stars – α$_{alkyl}$-relaxations (BDS); open circles – α$_{alkyl}$-relaxation (FSC); open squares – α$_{alkyl}$-relaxation (TMFSC); open pentagons – glass transition temperatures (DSC); open diamonds – α$_{alkyl}$-relaxation (TMDSC). The thermal relaxation rates were calculated by eqn. (4) for the DSC and FSC data. Dashed lines are the fits of the Arrhenius-equation (eqn. (2)) to the data. Solid lines are the fits of the VFT-equation (eqn. (3)) to the data. The data for **HAT5** and **HAT6** are taken from ref. 31 and ref. 32 respectively. Gray hexagons symbolize the dielectric γ-relaxation of polyethylene (**PE**) taken from ref. 19. The solid gray line denotes the dielectric α-relaxation of **PE** taken from ref. 50.

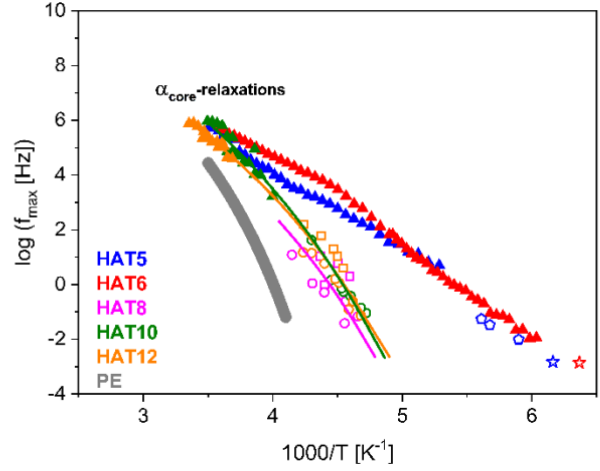

**Figure 6.** Relaxation map of **HATn** for the α$_{core}$-relaxation. Each **HATn** material is marked by a different color as indicated. Filled symbols denote the dielectric and open symbols denote the calorimetric data. Filled triangles – BDS; open circles – FSC; open squares – TMFSC; open pentagons – TMDSC; under helium purge. Solid lines are fits of the VFT-equation (eqn. (3)) to the data. **HAT5** and **HAT6** data are taken from ref. 31 and ref. 32 respectively. Solid gray line denotes the dielectric α-relaxation of **PE** taken from ref. 50.

The temperature dependence of the relaxation rates of the γ-relaxation follows the Arrhenius equation

$$log\, f_{max} = log\, f_\infty - \frac{E_A}{ln(10)\, RT}. \quad (2)$$

Here, $E_A$ symbolizes the activation energy. $f_\infty$ is the relaxation rate at infinite temperatures and R is the general gas constant. The activation energies for the γ-relaxation of the studied **HATn** materials have values in the range from 20 kJ/mol to 34 kJ/mol (see Table S1 in the SI). In references 30-34 for variety of CLCs including triphenylene-based DLCs, comparable values of the activation energies of the γ-relaxation (12-33 kJ/mol) are found. Moreover, data for the γ-process of **PE** is included in Figure 5. As a result, it is found that the activation energy of the γ-relaxation observed for **HATn** is comparable to that of the γ-process of polyethylene (**PE**).[19] Therefore, the γ-relaxation of **HATn** is attributed to localized fluctuations of the methylene groups occupying the intercolumnar area. This assignment is proofed for **HAT6** by neutron scattering investigations combined with BDS.[32]

Figure 7 shows the activation energy of the γ-process as a function of the alkyl chain length. $E_A$ first decreases with increasing alkyl chain length till ca. n=7 (hypothetical data point) and increases with further the increase of n again. Such a dependence can be discussed to result from a counterbalance of two competing effects. As discussed in ref. 51 for the vibrational density of states of the considered series of **HATn** materials, the order in the intercolumnar area depends on the length of the alkyl chain. It is reported that for **HAT5** and **HAT6** the intercolumnar area is more ordered or constrained than that of **HAT10** and **HAT12**. From these results, it is concluded that there might be a transition at n =8 with increasing alkyl chain length from a rigid (**HAT5** and **HAT6**) to a soft system (**HAT10** and **HAT12**). Based on the discussions in ref. 51, one might discuss the dependence of the activation energy of the γ-process on the alkyl chain length in the following way. **HAT6** has less ordered or less constrained intercolumnar area with a larger distance between the columns compared to **HAT5**. This might be considered as a partial release of the self-confinement superimposed to the methylene groups by the columns of the aromatic cores comparted to **HAT5**. This effect will lead to a decrease of the activation energy of the γ-process for **HAT6** compared to that of **HAT5**. With the further increase of the length of the alkyl chains the structure and the behavior of alkyl chain becomes more **PE**-like and also the γ-relaxation should become more **PE**-like. Therefore, the activation of the γ-process increases towards values which are characteristic for **PE**. Moreover, the discussion about the release of the self-confinement model and the change from a rigid to a softer system with increasing length of the substituents agrees with the arguments given in ref. 52 and 53.

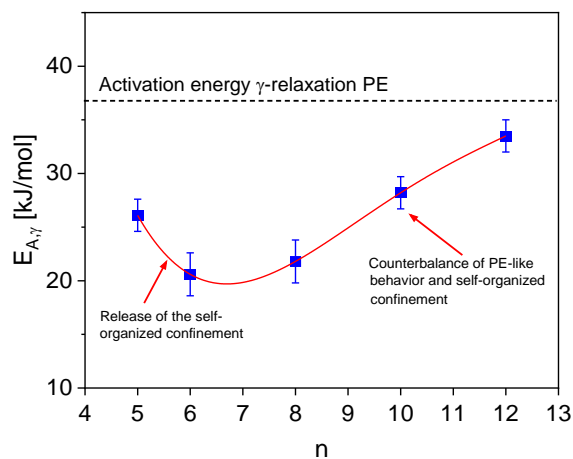

**Figure 7.** Activation energy of γ-relaxation *versus* the number of carbon atoms in the alkyl chain for **HATn**.

The dependence of the phase transition enthalpies from the Cry to the Col$_h$ phase on the length of the alkyl chain show a non-monotonous behavior (see Figure 5 in ref. 10). Also, the step-height of the specific heat capacity at the glass transition shows an anomaly in dependence on the chain length between n=6 and n=8 (see Figure 8). These results point also to a change in the behavior of HATs with short alkyl chains and longer ones, which is discussed in more detail in the following.

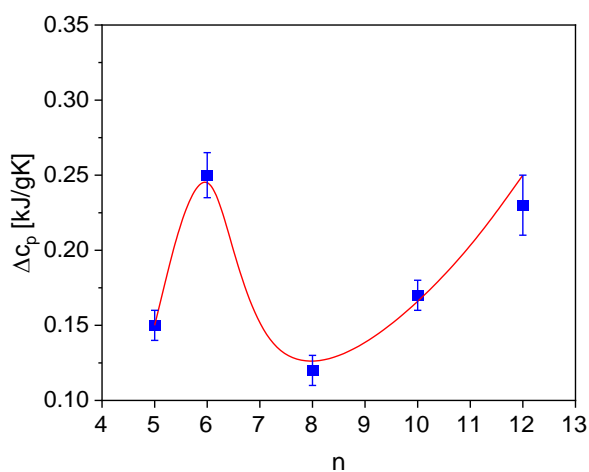

**Figure 8.** Change in the specific heat capacity versus the number of carbon atoms in the alkyl chain for **HATn**.

The temperature dependencies of the relaxation rates of both the α$_{core}$- for n>6 and the α$_{alkyl}$-processes for all samples in the Arrhenius diagram are curved, which is taken as an indication of glassy dynamics. For this reason, the temperature dependence of the relaxation rates of α$_{core}$- for n>6 and that of α$_{alkyl}$-relaxations for all **HATn** materials are described by the empirical Vogel/Fulcher/Tammann (VFT-) equation[54]

$$log\ f_{max} = log\ f_\infty - \frac{A}{T - T_0} \quad (3)$$

where A is a constant, and $T_0$ is the so-called Vogel or ideal glass transition temperature. The calculated VFT parameters are summarized in Table S1 in the SI. The temperature dependence of the relaxation rates of the α$_{alkyl}$-relaxation for all **HATn** and that of the α$_{core}$-relaxation for **HAT8**, **HAT10** and **HAT12** could be well-described by VFT-equation. However, for **HAT5** and **HAT6** the temperature dependence of the relaxation rates of α$_{core}$-relaxations is more complicated. For HAT5, it can be approximated by an Arrhenius equation but with a high apparent activation energy of 60 kJ/mol. For HAT6, it implies that the behavior changes with increasing temperature at ca. 220 K. As discussed in ref. 32, the alkyl chains filling in the intercolumnar area are frozen at lower temperatures and then they become more mobile at the glass transition temperature (206.6 K for **HAT5** and 214.0 K for **HAT6**). A further increase of the temperature, the passing of the glass transition makes the intercolumnar area softer. Therefore, the constrains of the frozen alkyl chains to the fluctuations of the cores decreases with increasing temperature. Such a change in the constrains on the fluctuations of the cores will affect the temperature dependence of the relaxation rates of α$_{core}$-process. For the alkyl chain length n>6, the distance of the columns is large enough that the constraint effect of the frozen alkyl chain on the dynamics of the cores becomes more minor.

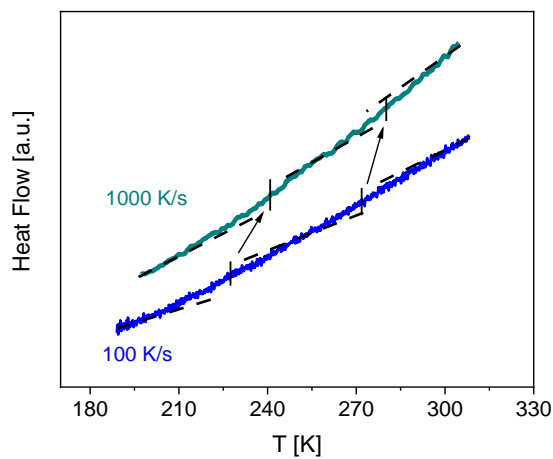

**Figure 9.** FSC heating thermograms of **HAT8** at the indicated frequencies. Solid vertical lines show the estimated glass transition temperatures. The curves are moved along the y-scale for the sake of clarity. The arrows indicated the shift of the glass transition temperature to higher temperatures with increasing frequency.

Studying the glassy dynamics by so-called advanced calorimetric techniques (e.g. FSC and SHS) has been demonstrated as a useful strategy complementing BDS investigations.[30-34] Therefore, the glassy dynamics obtained by BDS ($α_{core}$ and $α_{alkyl}$-processes) was further investigated by FSC measurements. As an example of the FSC investigations, Figure 9 shows heat flow curves for **HAT8** for heating rates in the range from 10 K/s to 10000 K/s. For **HAT10** and **HAT12**, the heat flow curves obtained by FSC measurements are given in Figure S2 and S3 respectively. Two separate step-like dependences at low and high temperatures were observed in the temperature dependence of the heat flow for all **HATn** compounds. These steps imply two glass transitions. Glass transition temperatures were estimated from the mid position of the step-like changes of the heat flow.

As discussed in literature[31-34], a thermal relaxation rate can be estimated from the FSC measurements using the heating rate $\dot{T}$ and the width of the glass transition region $\Delta T_g$ by [55,56]

$$f_{max} = \frac{\dot{T}}{2\pi a \Delta T_g} \quad (4)$$

$\Delta T_g$ was calculated by the difference between endset and onset temperatures of the step-like change of the heat flow at the glass transition. $a$ is a constant with a value of approximately one. Further, the relaxation rates for the conventional DSC were also calculated using eqn. (4) and added to Figure 5.

Figure 10 depicts the reversing heat capacity measured by SHS employing TMFSC for **HAT8** at 1 Hz and 11 Hz. For the analysis of the data, a sigmoidal function was fitted to the data. Then the first derivative of the fit function was taken with respect to temperature. Such a procedure will reduce the scatter of the data. This analysis results in two peaks, which shift to higher temperatures with increasing frequency as expected. The peak maximum temperatures in the derivative curve were taken as the glass transition temperatures for the different frequencies. The determined data are included to Figure 5 and Figure 6.

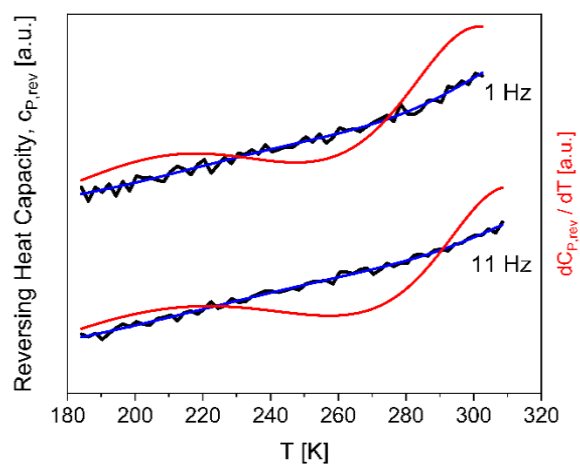

**Figure 10.** Reversing heat capacity curve obtained by TMFSC at a frequency of 1 Hz and 11 Hz for **HAT8**. The blue solid lines are sigmoidal fits to the data. The red curves are the first derivatives of the sigmoidal fits with respect to temperature. The frequency dependent glass transition temperatures are determined from the maximum temperatures of the peaks in the derivative.

In addition to the SHS study employing TMFSC, the SHS investigations employing TMDSC were also conducted to study the glassy dynamics at a lower frequency than those covered by TMFSC. Figure 11 gives the StepScan heat capacity versus temperature for the TMDSC measurements. The temperature dependence of the reversing heat capacity shows a step-like change with increasing temperature for each sample, which indicates a glass transition. (As the StepScan specific heat capacity can contain latent heat effects the peak in the heat flow curves close to the glass transition might result from enthalpy relaxation.) By taking the mid-step position of the temperature dependence of the reversing heat capacity the glass transition temperatures at a frequency of $1.6 \times 10^{-2}$ Hz were determined and added into Figure 5.

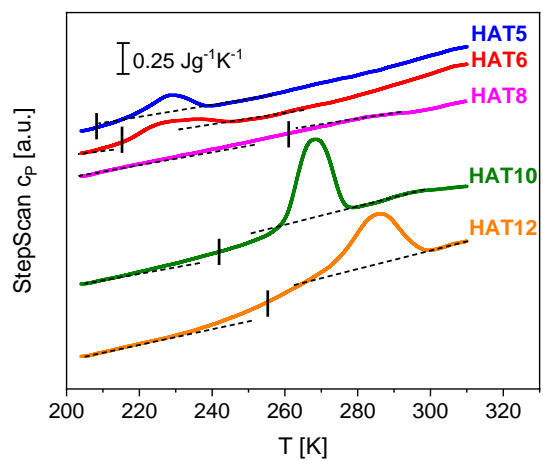

**Figure 11.** Temperature dependence of the reversing heat capacity at a frequency of $1.6 \times 10^{-2}$ Hz for **HATn**. Each **HATn** is marked by a different color as indicated. Black solid vertical lines indicate the determined glass transition temperatures. The curves are shifted along the y-scale for the sake of clarity (shifts indicated in the J/gK values in the brackets).

The temperature dependencies of the relaxation rates of both the dielectric and the calorimetric data for the $\alpha_{alkyl}$- and the $\alpha_{core}$-relaxation (n<6) processes can be described by a common VFT-dependence for each **HATn**, except that for $\alpha_{core}$-relaxation of **HAT5** and **HAT6** (see Figure 5 and Figure 6). As reported in references 31 and 32, for **HAT5** and **HAT6** the dielectric and the calorimetric data for the $\alpha_{core}$-process are in agreement and complementing each other. On the one hand, the temperature dependences of the relaxation rates of the $\alpha_{alkyl}$-relaxation of **HATn** are comparable to that of α-relaxation rates of **PE**. Among the studied systems especially the data for the $\alpha_{alkyl}$-relaxation rates of **HAT10** overlaps almost completely with the data for the α-relaxation rates of **PE** (see Figure 5). Thus, the $\alpha_{alkyl}$-relaxations of the **HATn** materials are attributed to **PE**-like fluctuations of the alkyl chains filling the intercolumnar area. On the other hand, the $\alpha_{core}$-relaxation of **HATn** are assigned to the fluctuations of the core including its rotational fluctuations around the column axis. Based on the discussions given in ref. 31, the assignment of the $\alpha_{core}$-relaxations to fluctuations of the core could not exclude the fluctuations of the alkyl chains surrounding the core. It might be that while the core rotates around the column axis, the methylene groups close to the core also fluctuates together with the motion of the cores.

The interpretations of the molecular origins of $\alpha_{alkyl}$- and the $\alpha_{core}$-processes are consistent with previous investigations on the molecular mobility in CLCs.[30,31,33] For comparison, a relaxation map showing the glassy dynamics of the different CLC systems (**SHU10**[31], **SHU12**[31], **KAL468**[33] and **LC536**[30]) is given in Figure S4. the temperature dependences of the $\alpha_{alkyl}$-relaxation rates of **HATn** are similar to that of **SHU10**, **SHU12**, **KAL468** and **LC536**. Moreover, the temperature dependences of the $\alpha_{core}$-relaxation rates of **HATn** have close resemblance to that of **SHU10**, **SHU12**, **KAL468** and **LC536**.

The temperature dependence of the $\alpha_{alkyl}$-relaxation of the **HATn** materials varies with the length of the alkyl chains, and it shifts to higher temperatures with increasing of the alkyl chains length except that of **HAT8** (see Figure 5). The temperature dependence of the $\alpha_{alkyl}$-relaxation of the **HAT8** were observed at higher temperatures than the temperatures that of the other **HATn** materials (n=5,6,10,12) detected. Such a non-monotonous dependence of the temperature dependence of the relaxation rates on the length of alkyl chains at n=8 is also observed for that of $\gamma$-process as discussed above. It is important to underline once again that the $\alpha_{alkyl}$-process takes place in the intercolumnar area, where nanophase separated alkyl chains are self-confined between the rigid columns of the cores. Such a **PE**-like glass transition within the alkyl nanodomains has been reported for several polymer series having alkyl groups in the side chains and discussed in the context of a hindered glass transition in self-assembled confinements.[11] Moreover, the size of the self-confinement (column-to-column distance for DLCs) increases with increasing alkyl chain length.[10,11] In the concept of a hindered glass transition[55], the **PE**-like glassy dynamics within the alkyl nanodomains changes depending on the size relation between the self-confinement size ($d_{self-confiment}$) and the size of characteristic length scale of the glass dynamics in the bulk **PE** ($\xi_{\alpha PE}$).[11] In this perspective, for short alkyl chain lengths (n=5,6) the $\xi_{\alpha PE}$ is larger than the $d_{self-confiment}$. In this case of so-called strong confinement ($\xi_{\alpha PE} \gg d_{self-confiment}$), and the temperature dependence of the $\alpha_{alkyl}$-relaxation of **HAT5** and **HAT6** shift to lower temperatures compared to that of the α-relaxation of **PE**. For long alkyl chain lengths (n>6) the $\xi_{\alpha PE}$ is bigger than or comparable to the $d_{self-confiment}$. In this case of so-called moderate confinement ($\xi_{\alpha PE} \geq d_{self-confiment}$), the self-confinement is weak and partially released and thus the temperature dependence of the $\alpha_{alkyl}$-relaxation of **HAT8**, **HAT10** and **HAT12** resembles more to that of α-relaxation of **PE** compared to **HAT5** and **HAT6**. This is discussed above in a similar way for the γ-process as a counterbalance of two competing effects of a **PE**-like behavior and self-organized confinement with a change in the counterbalance at ca. n=7. Moreover, similar discussion could be valid for the dependence of the core dynamics ($\alpha_{core}$-relaxation) on the length of the alkyl chains. Although the $\alpha_{core}$-relaxation in **HATn** is not directly related to the **PE**-like glassy dynamics in **HATn**, the cooperative fluctuations of the alkyl chains have presumably influence on the cooperative fluctuations of the core.

## 4. CONCLUSIONS

The phase behavior and molecular mobility of a homologous series of columnar liquid crystals, **HATn** (n=5,6,8,10,12), having the same aromatic triphenylene core with different lengths of the alkyl chains were investigated. The aim was to get an understanding of the effect of alkyl chain length on the molecular mobility of **HATn**. The phase behavior of **HATn** compounds was explored using conventional DSC. The results show that longer alkyl chains lead to decrease in both phase transition temperatures as already known from the literature.

Moreover, it is observed that the temperature range of the Col$_h$ mesophase which is the desired one for applications becomes wider with decreasing side chain length. Therefore, it is concluded that the stability of the columns formed by the molecules of the **HATn** weakens with increasing alkyl chain length, since the stability of the columns depends mainly on the core-core interactions and the flexibility of the alky chains as discussed.

The molecular mobility was studied in broad temperature and frequency range using BDS. The BDS investigation revealed three relaxation processes: a localized γ-relaxation as well as two glassy dynamics denoted as α$_{core}$- and α$_{alkyl}$-relaxation. All relaxation processes were observed in the Cry phase. The γ-relaxations of **HATn** are ascribed to the localized fluctuations of the methylene groups filling the space between the rigid columns built by the rigid triphenylene cores. The dependence of the estimated activation energy of the γ relaxation on the alkyl chain length showed that there is a transition at around n=8 from a rigid (n=5,6) to a softer (n=10,12) system with increasing chain length. It is proposed that **PE**-like behavior and self-organized confinement are two competing effects in these materials, and they are responsible for the change form a rigid to a soft system at ca. n=7.

Multiple glassy dynamics denoted as α$_{core}$- and α$_{alkyl}$-relaxation, were observed by both BDS and advance calorimetry including FSC, TMFSC and TMDSC. The α$_{alkyl}$-relaxation is ascribed to the **PE**-like fluctuations of the alkyl chains occupying the intercolumnar area, while the α$_{core}$-relaxations are assigned to the fluctuations of the core. It is concluded that the temperature dependence of these glassy dynamics varies with alkyl chain length due to the change in size of the self-organized confinement.

## ASSOCIATED CONTENT

Supporting Information is available.

The fit parameters of VFT and Arrhenius fits, the Arrhenius diagrams combining the data from BDS, FSC, SHS and DSC presented together with the literature data.

## AUTHOR INFORMATION

**Corresponding Authors**

Tel: +49 30/8104-3384; e-mail: Andreas.Schoenhals@bam.de.

**Notes**

The authors declare no competing financial interest.

## ACKNOWLEDGMENTS

The German Science Foundation (DFG) is acknowledged for financial support (SCHO 470/21-1, SCHO 470/25-1 and Project number 430146019) Moreover, this project was supported by the DFG within the Collaborative Research Initiative SFB 986 "Tailor-Made Multi-Scale Materials Systems" Project number 192346071.

**Authors Contributions:** AS and PH initiated and supervised the wok, AY conducted the calorimetric measurements, and analyzed all data. CK carried out the dielectric measurements. The manuscript was written by AY, PH and AS. All authors agree on the final version of the manuscript.